\documentclass{sig-alternate}
\usepackage[T1]{fontenc}
\usepackage[latin9]{inputenc}
\usepackage{array}
\usepackage{multirow}
\usepackage{amssymb}
\usepackage{graphicx}

\toappear{}

\newcommand{\enumspacing}{}

\makeatletter

\DeclareGraphicsExtensions{.ps,.eps,.epsi}

\providecommand{\tabularnewline}{\\}

\usepackage{url}

\DeclareMathOperator{\sgn}{sgn}

\makeatother

\begin{document}

\title{Is Somebody Watching Your Facebook Newsfeed?}

\author{
\alignauthor $^1$Shan-Hung Wu, $^2$Man-Ju Chou, $^3$Ming-Hung Wang,\\
$^4$Chun-Hsiung Tseng, $^2$Yuh-Jye Lee, and $^3$Kuan-Ta Chen\\
\vskip0.5pc
\affaddr{$^1$Dept. of Computer Science, National Tsing Hua University}\\
\affaddr{$^2$Dept. of Computer Science, National Taiwan University of Science and Technology}\\
\affaddr{$^2$Inst. of Information Science, Academia Sinica}\\
\affaddr{$^3$Dept. of Information Management, Nanhua University}\\
}

%\author{Anonymous}
\maketitle

\begin{abstract}

With the popularity of Social Networking Services (SNS), more and more sensitive
information are stored online and associated with SNS accounts. The obvious value of
SNS accounts motivates the \emph{usage stealing} problem---unauthorized, stealthy use
of SNS accounts on the devices owned/used by account owners without any technology
hacks. For example, anxious parents may use their kids' SNS accounts to inspect the
kids' social status; husbands/wives may use their spouses' SNS accounts to spot
possible affairs. Usage stealing could happen anywhere in any form, and seriously
invades the privacy of account owners. However, there is \emph{no any currently known
defense} against such usage stealing. To an SNS operator (e.g., Facebook Inc.), usage
stealing is hard to detect using traditional methods because such attackers come from
the same IP addresses/devices, use the same credentials, and share the same accounts as
the owners do.

In this paper, we propose a novel continuous authentication approach that analyzes user
browsing behavior to detect SNS usage stealing incidents. We use Facebook as a case
study and show that it is possible to detect such incidents by analyzing SNS browsing
behavior.  Our experiment results show that our proposal can achieve higher than 80\%
detection accuracy within 2 minutes, and higher than 90\% detection accuracy after 7
minutes of observation time.
\end{abstract}

\section{Introduction}

\label{sec:intro}

Many people use Social Networking Services (SNS, such as Facebook) daily, and have
associated a lot of personal and sensitive information with their SNS accounts. These
information generally include friend lists, feeds from their friends, non-public
posts/photos, private interactions with others (such as chats and messages), purchased
apps/items, etc., and its obvious value makes the SNS accounts one of the most targeted
online resources by hackers to steal. To protect user privacy, SNS sites today have
done a lot to prevent account stealing. For example, Facebook records the regular IP
addresses and devices used by each account. If an account logs in with an unusual IP
address or device, the account is prompted to either answer some secret
questions~\cite{constine:friends:10} or enter a security code sent to the account
owner\textquoteright{}s mobile device~\cite{constine:phone:12} in order to verify if
the login is authentic. Facebook also allows users to report account stealing manually
if they suspect that their accounts have been stolen.

Despite of all the efforts to prevent account stealing, user privacy can also be
compromised by another form of breach called \emph{usage stealing}---unauthorized,
stealthy use of SNS accounts on the devices owned/used by account owners without any
technology hacks. Usage stealing could happen anywhere in any form. For example,
anxious parents may use their kids' SNS accounts to inspect the kids' social status;
husbands/wives may use their spouses' SNS accounts to spot possible affairs. Similarly,
colleges, supervisors, friends, or siblings, just to name a few, may also have their
own motives to use acquaintances' accounts for different reasons.

Usage stealing is common in practice due to the following reasons. First, when using
their own computers, people tend to choose ``yes'' when the browsers ask whether they
would like to save their (SNS) passwords for automatic logins in the future. This is
especially true when users are using their mobile devices because it is cumbersome to
input passwords~\cite{credant:phone,mah:store:2011}. Mobile devices also ease the usage
stealing in other aspects, in that they can be physically accessed by acquaintances or
strangers~\cite{yu:lost:2012}, and that most of them are not locked by
PINs~\cite{hansberry:pinlock:2011}. In addition, many SNS sites use cookies to save the
trouble of future account authentications within a short time. For example, once logged
into Facebook, a user need not login again during at most 60 following
days~\cite{Facebook:offline}. From the above, if someone (mostly an acquaintance) can
access the computer or mobile devices of an SNS user, it is likely that the person,
\emph{without the need of technical background}, can peep the information associated
with the SNS account.

However, there is \emph{no any currently known defense against such usage stealing}. To
an SNS site, usage stealing is hard to detect using traditional methods because the
attackers come from the same IP addresses/devices, use the same credentials, and share
the same accounts as the owners do. Moreover, because users normally do not see the
logs, victims can hardly sense and report the stealthy usage.

\textbf{Contributions.} In this paper, we identify the usage stealing problem in SNS
and propose a novel continuous authentication
approach~\cite{pusara:user:04,shepherd:continuous:95} that analyzes users' browsing
behavior to detect usage stealing incidents. We use Facebook as a case study, and show
that it is possible to detect such incidents by analyzing their browsing behavior on
the SNS sites, namely, clicks on newsfeeds\footnote{A Facebook newsfeed, which locates
at the center column of one's home page, is a constantly updated list summarizing the
status of people that one follows on Facebook.}, friend lists, profiles, likes,
messages, photos/videos, and comments. Our user study shows that our proposed scheme
can achieve above 80\% accuracy with a high confidence within 2 minutes, and higher
than 90\% accuracy after 7 minutes of observation time.

\textbf{Deployment.} Our detection approach is designed to run on SNS servers and to
serve as \emph{the first line of defense} against account usage stealing. The
deployment is straightforward: an SNS server collects the behavior of an account's
session and feeds it into a detection model \emph{in real time}. The model determines
whether the user of the session is suspicious, and if so, the SNS server can either 1)
trigger more sophisticated analysis/monitoring, and/or 2) challenge the session user
immediately by secret questions or via a second channel such as mobile phone
authentication~\cite{constine:phone:12}. Note that since we are at the first line of defense, there is
no need for an 100\% accurate detection model, rather, a reasonable detection power is
sufficient and the key is a prompt and efficient detection. Also, please note that the
proposed methodology is neither tied to a specific SNS site nor to a certain learning
technique as it is based on the standard supervised learning framework. For example,
while we adopt the smooth SVM~\cite{lee2001ssvm} as the detection model in this paper,
the service operators (e.g., Facebook Inc.) may choose the asymmetric
SVM~\cite{wu:asvm:08} or similar methods if they wish to further reduce the false
positive rates.

\textbf{Implications.} We believe that the usage stealing problem, while not being well
studied so far, will be much more critical in the future, as people put more and more
sensitive information online. In fact, this problem may not only occur in the social
services, but also online email services such as Gmail and Outlook.com, time management
services such as Google Calender and Remember The Milk, photo album services such as
Instagram, and much more. Except asking users to repeatedly authenticate themselves
(practically prohibited by usability issues), continuous authentication seems to be the
only feasible solution for attacks of this kind.

The rest of this paper is organized as follows. We review the related work in Section
\ref{sec:rel}. Section \ref{sec:rationale} discusses the rationale behind detecting
usage stealing based on browsing behavior. In Section \ref{sec:behavior}, we describe
our user study on Facebook and analyze the users' behavior. Section \ref{sec:detection}
elaborates our detection methodology. We evaluate the performance of our scheme in
Section \ref{sec:eval}, and analyze the security issues in Section \ref{sec:security}.
Finally, Section \ref{sec:concl} concludes the paper.

\section{Related Work}

\label{sec:rel}

In this section, we review existing studies on the privacy issues on SNS and continuous
authentication.

\textbf{SNS Privacy.} Privacy is always a concern for SNS users. Many efforts have been
devoted to protect user privacy. He et al., Zheleva et al., and Tang et
al.~\cite{He:privacy:06,Tang:name:11,Zheleva:illusion:09} observe a privacy hole that
attackers can infer private information (such as sexual orientation) of a user from
his/her public SNS records/activities. Felt et al. and Wishart et
al.~\cite{Felt:privacy:08,Wishart:butler:10} prevent privacy leaks from SNS developer
APIs and from the software based on them. Mahmood et al.~\cite{Mahmood:deactivate:12}
focus on another type of privacy attacks called the frequent account deactivation.
Meanwhile, in the industry, Facebook takes a vector of measures to protect user
privacy. For example, it provides an official page~\cite{Facebook:security} to educate
users about the correct privacy and security settings, and records the IP addresses,
web browsers, and devices used by each account~\cite{Facebook:policy:09}. If an account
logs-in with unknown records, Facebook will challenge the user either by secret
questions~\cite{constine:friends:10} or via mobile phone
authentication~\cite{constine:phone:12}.  Facebook also allows users to report account
stealing incidents manually.

However, none of the above attempts can protect user privacy when the attackers sneak
in using the \emph{same devices} owned by the victims. Since the passwords,
credentials, and cookies are usually stored in users' devices to avoid repeated account
authentication~\cite{Facebook:offline,mah:store:2011,credant:phone}, attackers who have
physical access to these devices can easily bypass all the above detection schemes and
obtain the sensitive information associated with the SNS accounts.

\textbf{Continuous Authentication.} Continuous authentication is an implicit, automatic
re-authentication method that analyzes the follow-up user actions after his/her initial
authentication to make sure if the user is still authentic. The actions can be keyboard
typing behavior~\cite{shepherd:continuous:95}, mouse movements~\cite{pusara:user:04},
operations on mobile devices~\cite{yazji:implicit:09}, facial characteristics (if a
webcam is available)~\cite{niinuma:continuous:10}, or any other soft biometric
traits~\cite{niinuma:soft:10,yap:physical:08}.

However, the above analyses are per-person-based; that is, a detection model is
required for each user. This may be cost-prohibitive on SNS servers given that an SNS
site usually have more than millions of users\footnote{For example, Facebook has more
than a billion monthly active users as of December 2012.}. The continuous
authentication method proposed in this paper analyzes web browsing behavior performed
by only three predefined user groups. The detection model is universal to \emph{all}
users and it introduces low data collection and computation overhead. Another advantage
of our proposal is that the scheme can be applied to a new account whose associated
biometric behavior is not yet clear. Note that our proposal is not a replacement for
existing continuous authentication approaches. Rather, it can serve a low-cost filter
for suspicious accounts, with which the servers can trigger more sophisticated,
personalized analysis whenever necessary.

\section{Rationale behind Our Detection Approach}

\label{sec:rationale}

Nowadays, an SNS service such as Facebook is not merely a place for people to maintain
their friend lists. They are more like a platform where people engage various social
activities, such as posting own status, reading others\textquoteright{} comments on
news, chatting, and meeting new people, etc.
Studies~\cite{Hampton:fbuser:11,joinson:looking:08} show that there is \emph{no
typical} user behavioral pattern on a complicated, open platform like Facebook, as
every single user seems to have his/her own behavioral tendency on an SNS service. For
example, some people tend to fulfill their desire on self-presentation, so they spend
most time on sharing their own latest status and posting the latest photos/events. In
the meantime, some people may manage to engage new friends online; some chat with
familiar friends; some spend time discovering new social games; and some others like to
stalk certain other users.

Given the diversity in user behavioral patterns determined by users\textquoteright{}
personal characteristics and social status, it is hard to profile every
user\textquoteright{}s browsing behavior when they are using an SNS service. However,
we argue that users would normally exhibit significantly different behavior when they
are browsing their own and others' pages.

In the context of usage stealing, each user can have one of the three following roles
when using an SNS service: 1) \emph{owner}, when he/she is using his/her own account;
2) \emph{acquaintance} (as a stalker), when he/she is using the account of someone
he/she knows; and 3) \emph{stranger} (as a stalker), when he/she is using the account
of an unknown person. Intuitively, when checking the Facebook newsfeed as the owner, a
user would focus more on the latest information of friends and use the ``like'' or
``share'' function to interact with others. On the other hand, when browsing a newsfeed
as a stalker (either an acquaintance or a stranger), the user may be interested in
earlier information that is more interesting to the stalker. He/she may not interact
with others because he/she does not want the owner to discover the stealthy usage
later. In summary, we believe that users would normally behave differently at different
roles because
\begin{itemize}
\item The way people treat familiar information (or information from familiar
friends) would be different than the way they treat unfamiliar information;

\item People at different roles would have different intentions;

\item In order not to be found by the account owners, people as the stalkers may behave
differently with the time pressure.
\end{itemize}
We call the above differences in users\textquoteright{} browsing behavior as the
\emph{role-driven behavioral diversity}. We conjecture that Facebook users, as well as
users of the other SNS services, possess such diversity, and this serves as the main
rationale behind our detection scheme.

In the following, we shall prove that the role-driven behavioral diversity indeed
exists using the datasets we collect in a user study on Facebook (Section
\ref{sec:behavior}) and then show that our detection scheme can rely on this property
to classify account owners from stalkers (Section \ref{sec:detection}).

\section{Facebook User Behavior}

\label{sec:behavior}

We use Facebook as a case study on users' role-driven behavior.

\subsection{Data Collection}

\label{sec-behavior-collect}

To capture the role-driven behavioral diversity, we hire a number of Facebook users to
be our subjects and design experiments in which subjects browse Facebook newsfeed at
different roles. In other words, we ask each subject to browse 1) his/her own newsfeed,
2) his/her friend\textquoteright{}s newsfeed, and 3) a stranger\textquoteright{}s
newsfeed.

To conduct the experiment, we hire \emph{pairs} of subjects from an one-million-user
Internet community. Each pair of subjects must be with at least one of the following
relationships: friends, family members, colleagues, classmates, and couples. Each
subject is paid 10 USD and we get the subject\textquoteright{}s permission to record
all actions (e.g., clicks, typing, page views, etc.) he/she performs when browsing a
newsfeed. A subject is hired only if he/she is an active Facebook user---the subject
must have more than 50 friends and consistently stay on Facebook longer than 4 hours
per week.

\begin{figure}[t]
\begin{centering}
\includegraphics[width=8cm]{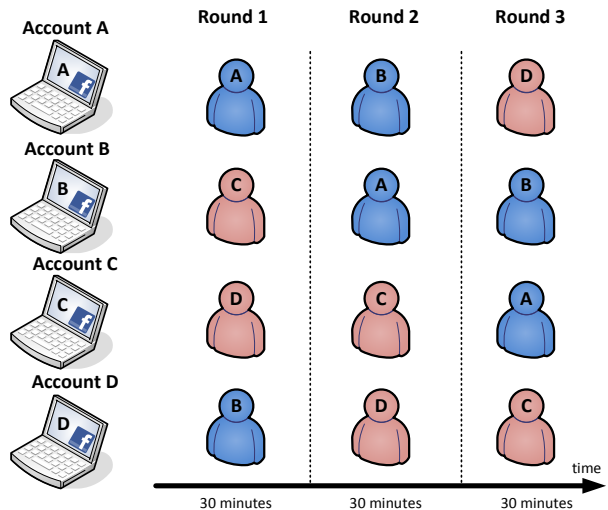}
\par\end{centering}

\caption{\label{fig:experiment}$(A,B)$ and $(C,D)$ are pairs of acquaintances. Each
experiment comprises 3 rounds. In each round, each subject is assigned to an account
\emph{randomly}; overall, each subject is guaranteed to browse his/her own account, the
account of acquaintance, and an account of stranger in the three rounds with a
randomized order.}
\end{figure}

Each experiment comprises 3 rounds. In each round, a subject is asked to browse the
newsfeed of an account (either of his/her, his/her friend\textquoteright{}s, or a
stranger\textquoteright{}s own) for 30 minutes. The subjects and accounts are paired
\emph{randomly}, but each subject is guaranteed to play all the 3 roles in the 3
rounds, as shown in Figure~\ref{fig:experiment}. During each round, the users can
perform any action they like (e.g., uploading photos, leaving comments, etc.), but not
including sabotage activities (e.g., changing the account password). Actions to follow
external links (such as videos) are allowed, but the subjects are asked not to stay
with external content for longer than 1 minute each time.

\begin{table}
\begin{centering}
\begin{tabular}{l|l}
\textbf{\small{Property\quad{}\quad{}\quad{}}} & \textbf{\small{\quad{}Value}}\tabularnewline
\hline
\hline
{\small{\# experiments}} & {\small{28}}\tabularnewline
{\small{Total time}} & {\small{9302 min}}\tabularnewline
{\small{\# subjects}} & {\small{112}}\tabularnewline
{\small{\# male subjects}} & {\small{56}}\tabularnewline
{\small{\# female subjects}} & {\small{44}}\tabularnewline
{\small{\# sessions}} & {\small{278}}\tabularnewline
{\small{\# self-usage}} & {\small{100}}\tabularnewline
{\small{\# acquaintance-usage}} & {\small{81}}\tabularnewline
{\small{\# stranger-usage}} & {\small{97}}\tabularnewline
{\small{Avg. session length}} & {\small{30 min}}\tabularnewline
{\small{Avg. action rate}} & {\small{3.0 action/min}}\tabularnewline
{\small{Avg. page switching rate}} & {\small{0.7 page/min}}\tabularnewline
\end{tabular}
\par\end{centering}

\caption{\label{tab:dataset}A summary of the experiments and raw dataset.}
\end{table}

In order to capture the subjects' activities performed on Facebook, we use Fiddler, a
free Web debugging proxy, to monitor all the HTTP/HTTPS GET/POST requests issued by a
browser. By parsing the HTTP/HTTPS request logs, we are able to capture every action
performed by a user (to be explained later). Table~\ref{tab:dataset} summarizes the
experiments and the raw dataset we collect. Note that, although we have administrators
overseeing the experiment process, some subjects do not follow the contract to browse
the given Facebook newsfeeds. Out of 311 sessions, we remove 33 ``noisy sessions'' of
those subjects who are obviously not focusing on the newsfeeds, i.e., sessions with any
idle, non-active period longer than 5 minutes.

\begin{table}
\begin{centering}
\begin{tabular}{l|cc}
\textbf{\small{\quad{}\quad{}Actions }} & \textbf{\small{Interactive }} & \textbf{\small{}}%
\begin{tabular}{c}
\textbf{\small{Page-}}\tabularnewline
\textbf{\small{Switching}}\tabularnewline
\end{tabular}\textbf{\small{ }}\tabularnewline
\hline
\hline
{\small{Expand Comments }} & {\small{$\surd$ }} & \tabularnewline
{\small{Likes }} & {\small{$\surd$ }} & \tabularnewline
{\small{View Cards }} & {\small{$\surd$ }} & \tabularnewline
{\small{View Likes }} & {\small{$\surd$ }} & \tabularnewline
{\small{View Messages }} & {\small{$\surd$ }} & \tabularnewline
{\small{View Photos }} & {\small{$\surd$ }} & \tabularnewline
{\small{To Friend List Page }} & {\small{$\surd$ }} & {\small{$\surd$ }}\tabularnewline
{\small{To Note Page }} & {\small{$\surd$ }} & {\small{$\surd$ }}\tabularnewline
{\small{To Photo Page }} & {\small{$\surd$ }} & {\small{$\surd$ }}\tabularnewline
{\small{To Wall Page }} & {\small{$\surd$ }} & {\small{$\surd$ }}\tabularnewline
{\small{To Fan Page }} &  & {\small{$\surd$ }}\tabularnewline
{\small{To Feed Page }} &  & {\small{$\surd$ }}\tabularnewline
{\small{To Group Page }} &  & {\small{$\surd$ }}\tabularnewline
{\small{To Message Page }} &  & {\small{$\surd$ }}\tabularnewline
{\small{Add Comments }} &  & \tabularnewline
{\small{Delete Comments }} &  & \tabularnewline
{\small{Click Hyper-links }} &  & \tabularnewline
{\small{Expand Page }} &  & \tabularnewline
\end{tabular}
\par\end{centering}

\caption{\label{tab:action-type}18 types of common user actions we collected on Facebook.}
\end{table}

We categorize the common user activities on Fcebook into 18 different actions, as shown
in Table~\ref{tab:action-type}. Among these actions we can identify 2 groups: 1)
\emph{interactive actions}: those by which the user interacts with a certain person;
and 2) \emph{page-switching actions}: those lead the browser to switch to another
Facebook page. By parsing the HTTP/HTTPS request logs, we obtain a chronological action
list for each session, as exemplified in Table~\ref{tab:action-list}. Each record in the
list contains the action name, the occurrence time stamp, and, if the action is
interactive, the target person the user interacts with. Based on the account
owner's profile and friend list, we annotate the target person with either the ``account
owner,'' ``friend,'' or ``non-friend.''

\begin{table}
\begin{centering}
\begin{tabular}{l|l|l}
\textbf{\small{Time stamp }} & \textbf{\small{Action }} & \textbf{\small{Target Person }}\tabularnewline
\hline
\hline
{\small{\texttt{1345837539249.47} }} & {\small{Likes }} & {\small{Friend A }}\tabularnewline
{\small{\texttt{1345837568519.15} }} & {\small{View Cards }} & {\small{Account Owner}}\tabularnewline
{\small{\texttt{1345837586398.26}}} & {\small{Add Comment }} & {\small{Friend A}}\tabularnewline
{\small{\texttt{1345837732512.73} }} & {\small{Group page }} & \tabularnewline
{\small{\texttt{1345837756445.03} }} & {\small{Likes }} & {\small{Friend B}}\tabularnewline
{\small{\texttt{1345837770260.55} }} & {\small{View Cards }} & {\small{Non-Friend C}}\tabularnewline
{\small{\texttt{1345837773293.04} }} & {\small{View Message }} & {\small{Friend A}}\tabularnewline
{\small{\texttt{1345837828598.01} }} & {\small{Likes }} & {\small{Non-Friend C }}\tabularnewline
{\small{\texttt{1345837875240.45}}} & {\small{Expand Page }} & \tabularnewline
\end{tabular}
\par\end{centering}

\caption{\label{tab:action-list}An exemplary list of the collected actions.}
\end{table}

\subsection{Defining Features}

\label{sec:behavior-feature}

Next, we define a number of features and obtain their values for every session we
collect so that the machine learning algorithms can be applied.
%Extracting appropriate features from raw data might be the most important step in
%applying machine learning algorithms.
Even we have a perfect learning algorithm, without features that encode information
about who is controlling a session, the algorithm will have no way to distinguish the
account owners from stalkers. How to define features that we need is a key issue in
this work, and is usually challenging because it requires insights, domain knowledge,
creativity, and even ``black arts''~\cite{knowML}.

We interview heavy users of Facebook about their regular usage patterns and the ways
they discover and explore interesting information. Based on the results, we define 139
features. All the features of a particular session can be extracted from the session's
action list (see Table~\ref{tab:action-list}). Our features can be basically summarized
as follows:

\begin{enumerate}
\enumspacing

\item \texttt{f.<action>}: the frequency of a certain action (per minute). The
\texttt{<action>} can be any action defined in Table~\ref{tab:action-type}. We also
keep \texttt{f.acts} and \texttt{f.acts.excluding.\\page.expand}, the frequencies of
all actions and all actions except the ``expand page\footnote{In Facebook, some pages
(e.g., newsfeeds, walls, etc.) and page items (e.g., comments, notifications lists,
etc.) are expandable. For clarity, these pages/items show earlier/detailed information
only upon expansion.}'' action respectively. The reason we capture the latter feature
is that we want to determine how much a user really does in addition to merely browsing
pages.

\item \texttt{f.<target\_type>.<action>}: the frequency of a certain action targeting a
certain target user type. The \texttt{<action>} is an interactive action in
Table~\ref{tab:action-type} and \texttt{<target\_type>} can be \texttt{self} (if the
target person is the account owner), \texttt{friend} (if the target person is a friend
of the account owner), or \texttt{nonfriend} (if the target person is not a friend).

\item \texttt{b.<xxx>}: the binary version of all the above features; i.e.,
\texttt{b.<xxx>}$=1$ iff \texttt{f.<xxx>} is greater than $0$. For example,
\texttt{b.<action>} denotes whether or not a certain action occurs during the session.

\item \texttt{f.act.<target\_type>}: the frequency of all interactive actions performed
on a certain target user type.

\item \texttt{ts.page.<page\_type>}: the time the session user spends on a certain page
type. The \texttt{<page\_type>} can be \texttt{feed} (the account's newsfeed),
\texttt{msg} (the account's message box), \texttt{self} (pages, such as the wall/friend
list/note/photos, of the account owner), \texttt{friend} (pages of friends),
\texttt{nonfriend} (pages of non-friends), or \texttt{public} (fans or groups pages).

\item \texttt{f.act.page.<page\_type>}: the frequency of all actions performed on a
certain page type. We also keep \texttt{f.act.\\expand.page.<page\_type>} and
\texttt{f.act.non.expand.page.\\<page\_type>}, the frequencies of the ``expand page''
action and all other actions performed on a certain page type respectively.

\item \texttt{n.act.person}: the number of target people the user interacts
with during the session.

\item \texttt{n.act.person.<statistics>}: the statistics of the counts of visits to
different users' pages during the session. The \texttt{<statistics>} include
\texttt{mean}, \texttt{standard\_deviation}, \texttt{median}, and \texttt{maximum}. For
example, suppose that the user visits his/her own pages $1$ time, friend A's pages $3$
times, friend B's pages $1$ time, and non-friend C's pages $2$ times, then we obtain
\texttt{mean}$=1.75$, \texttt{standard\_\\deviation}$=0.96$, \texttt{median}$=1.5$, and
\texttt{maximum}$=3$. The reason we capture these features is that we want to determine
whether a user pays attention to any specific person.

\end{enumerate}
After extracting the features for each session, we obtain a dataset ready for analysis.
Each session is labeled with either ``owner,'' ``acquaintance,'' or ``stranger,''
depending on the user's role for the session in the experiment.

\subsection{Role-Driven Behavioral Diversity}
\label{sec:behavior-diversity}

To justify the existence of the role-driven behavioral diversity between the account
owners, acquaintances, and strangers, we carry out some analysis of the user behavior
performed at different roles. Our observations are summarized as follows.

\begin{figure*}
\begin{centering}
\includegraphics[width=17cm]{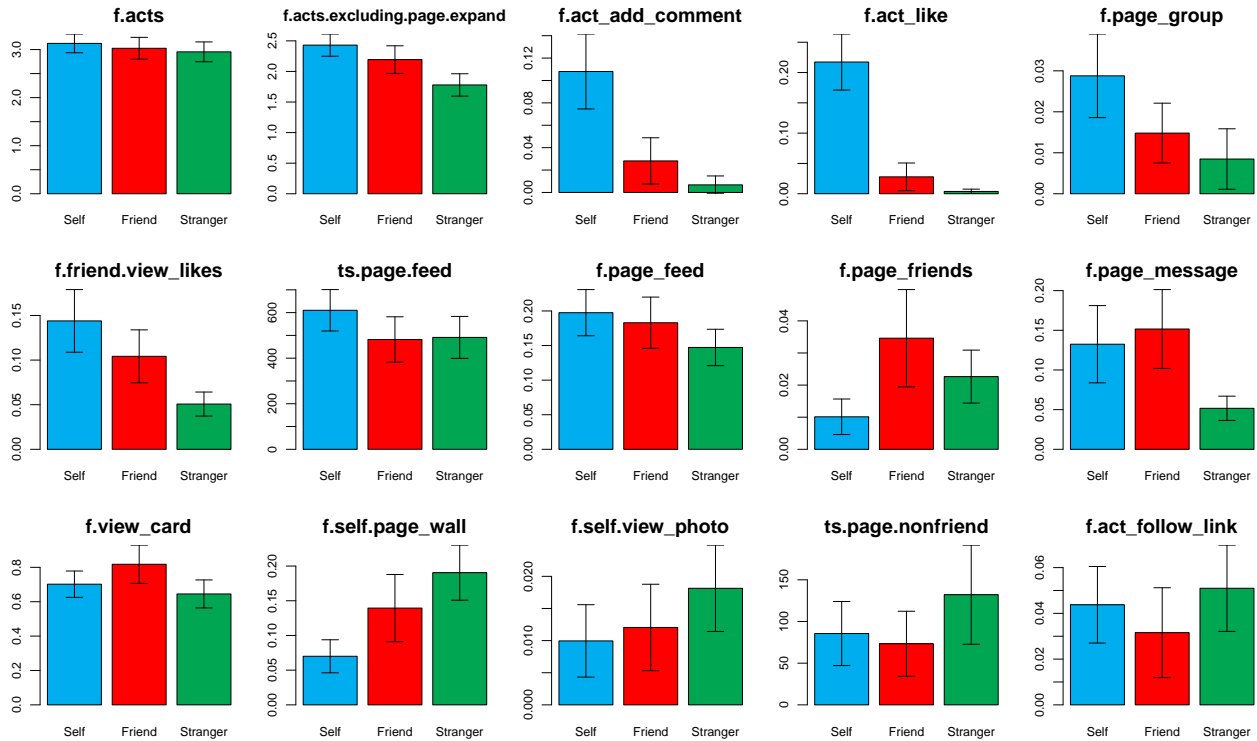}
\par\end{centering}

\caption{\label{fig:evidence}The evidence of role-driven behavioral diversity.}
\end{figure*}

\textbf{General Diversity.} As shown in Figure \ref{fig:evidence}(a), all sessions
controlled by the three user roles have similar values in \texttt{f.acts}. However, in
\texttt{f.acts.excluding.page.expand} (Figure \ref{fig:evidence}(b)), the sessions
controlled by the account owners exhibit higher values than those by the acquaintances,
which in term give higher values than those by strangers. This implies that the
acquaintances/strangers usually pay more attention to reading/searching for interesting
information. They also care more about earlier information, as the content hidden
by expandable pages/items by default are older.

The sessions used by acquaintances/strangers also yield much lower values in
\texttt{f.act\_add\_comment} (Figure \ref{fig:evidence}(c)) and \texttt{f.act\_like}
(Figure \ref{fig:evidence}(d)) than those by the account owners. The reason is obvious:
Acquaintances/strangers normally do not want to leave a clue of their peeping behavior.

\textbf{What Stalkers Do Not Care.} Although the acquaintances/strangers expand pages
more frequently (Figure \ref{fig:evidence}(b)), they do not expand the comment lists as
commonly as the account owners do. This is because the acquaintances/strangers may not
know the people leaving the comments, therefore showing less interest to them. Due to
similar reasons, acquaintances/strangers also express less interests in the fans/
groups pages (Figure \ref{fig:evidence}(e)) and who likes a post (Figure
\ref{fig:evidence}(f)); in addition, they also spend less time in the accounts'
newsfeeds (Figure \ref{fig:evidence}(g)). In particular, strangers tend to ignore most
of the newsfeeds because they generally do not know people whose information appear in
the feeds (Figure \ref{fig:evidence}(h)).

\textbf{What Acquainted Stalkers Care.} As compared with the other roles, the
acquaintances pay more attention to the accounts' friend lists (Figure
\ref{fig:evidence}(i)). This is because an acquaintance may know the account owner's
friends and be curious about these friends' status (especially the status of those
people who are not currently a friend of the acquaintance). The acquaintances are
generally most interested in the message boxes (Figure \ref{fig:evidence}(j)) and the
profile cards of the accounts' friends (Figure \ref{fig:evidence}(k)) due to similar
reasons.

\textbf{What Stranger Stalkers Care.} Interestingly, the account owners' profiles
(Figure \ref{fig:evidence}(l)) and photos (Figure \ref{fig:evidence}(m)) are most
viewed by strangers rather than the account owners' friends or the owners themselves.
This is because the strangers do not know the account owners, so they are usually
curious about who the owners are and how they look like. The strangers are also less
affected by the account owners' social relationship. For example, they are more willing
to check out non-friends (Figure \ref{fig:evidence}(n)) and external links (Figure
\ref{fig:evidence}(o)).

We believe that the above findings suffice to prove the existence of the role-driven
behavioral diversity. Next, we show how this diversity can be further utilized to
implement a low-cost detector for usage stealing.

\section{Detection Scheme}

\label{sec:detection}

This section introduces a scheme for detecting the usage stealing on SNS sites.

In our dataset, each session is labeled with either ``account owner,''
``acquaintance,'' or ``stranger.'' Since our goal is to distinguish stalkers from the
account owners, in the following, we replace the ``acquaintance'' and ``stranger''
labels with a single ``stalker'' label.

\begin{figure}[t]
\begin{centering}
\includegraphics[width=\columnwidth]{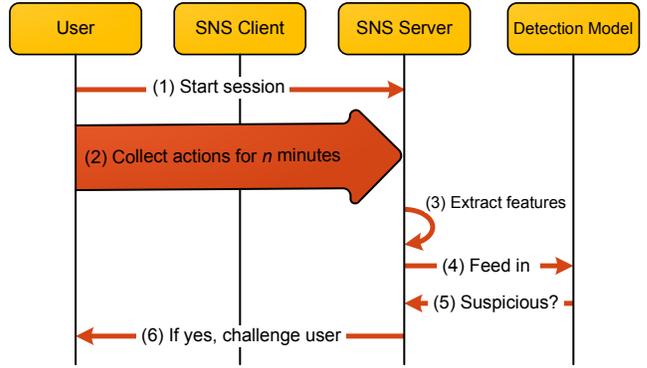}
\par\end{centering}

\caption{\label{fig:detect-scheme}The flow chart of our detection scheme.}
\end{figure}

Figure \ref{fig:detect-scheme} gives an overview of our detection scheme. After a user
starts a session (by either logging-in newly or using existing authentication cookies,
Step 1), the SNS server monitors and collects a list of actions performed by the user
for an \emph{observation period} of $n$ minutes, where $n$ is a configurable parameter
(Step 2). After the observation period, the SNS server extracts the features of the
monitored session based on the recorded action list (Step 3), where the features are
defined in Section~\ref{sec:behavior-feature}. It then feeds the session features into
a detection model (Step 4), which determines whether the session owner is suspicious by
predicting the label of the session (Step 5). If the predicted label is ``stalker,''
the SNS server can challenge the user by secret questions or via a second channel such
as mobile phone authentication (Step 6). Alternatively, the server can trigger a more
sophisticated (but costly) detection scheme.

Note that the scheme has low runtime cost on an SNS server because it requires only
\emph{one} detection model for \emph{all} SNS users, taking the advantage of the
role-driven behavioral diversity. Also note that, although we employ a two-class
detection model to distinguish stalkers from the account owners, the scheme can be
readily extended to identify the account owners, acquaintances, and strangers if a
multi-class detection model is adopted.

We obtain our detection model by training it using the labeled sessions we collected.
Clearly, the effectiveness of our detection scheme largely depends on the quality of
predictions made by the detection model. In order to obtain high-quality predictions,
we take rigorous steps in training the model, as summarized below.

\subsection{Model Development}

\label{sec:detection-model}

For ease of numeric operations, we treat the labels ``stalker'' and ``account owner''
as $1$ and $-1$ respectively. Given a training dataset $D$ of size $n$,
$D=\{(\mathbf{x}_{1},y_{1}),(\mathbf{x}_{2},y_{2}),\cdots,(\mathbf{x}_{n},y_{n})\}$,
where $\mathbf{x}_{i}\in\mathbb{R}^{d}$ is a labeled instance (i.e., session) with $d$
features and $y_{i}\in\{1,-1\}$ is the corresponding label, our goal is to obtain a
function $f:\mathbb{R}^{d}\rightarrow\mathbb{R}$ such that given a new instance
$\mathbf{x}'$ with an unknown label $y'$, we have $f(\mathbf{x}')>0$ iff $y'=1$. The
function $f$ is the detection model in our scheme. In Figure
\ref{fig:detect-scheme}(4), the SNS server feeds the session $\mathbf{x}'$ into $f$. In
Figure \ref{fig:detect-scheme}(5) the SNS server gets $f(\mathbf{x}')$ and determines
whether the session is suspicious by $\sgn(f(\mathbf{x}'))$.

\begin{figure}[t]
\centering{}\includegraphics[scale=0.28]{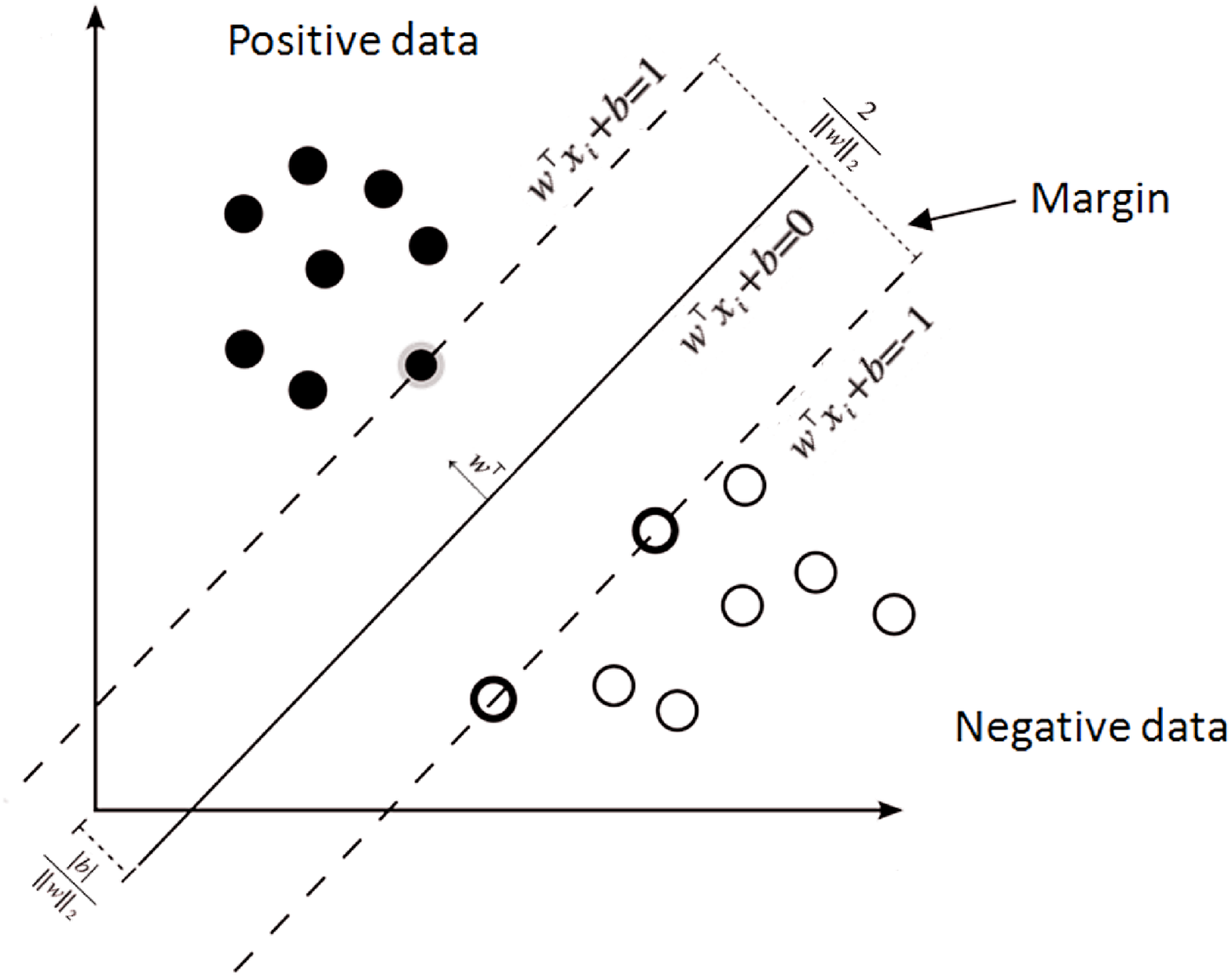} \caption{\label{fig:svm}The separating hyperplane and margin of SVM.}
\end{figure}

We obtain $f$ based on the \emph{Support Vector Machine} (SVM) technique.
The SVM is a very popular and promising machine learning algorithm
for binary classification problems. The idea is to let $f$ be a linear
function, i.e., $f(\mathbf{x})=\mathbf{w}^{\top}\mathbf{x}+b$ for
some $\mathbf{w}\in\mathbb{R}^{d}$ and $b\in\mathbb{R}$, and then
find $\mathbf{w}$ and $b$ such that the hyperplane $\{\mathbf{x}:\mathbf{w}^{\top}\mathbf{x}+b=0\}$
separates the positive and negative instances in $D$ while leaving
the largest ``margin'' between $\{\mathbf{x}:\mathbf{w}^{\top}\mathbf{x}+b=1\}$
and $\{\mathbf{x}:\mathbf{w}^{\top}\mathbf{x}+b=-1\}$, i.e., $\mathbf{w}^{\top}\mathbf{x}_{i}+b\geq1$
for all $(\mathbf{x}_{i},y_{i})\in D\,\wedge\, y_{i}=1$ and $\mathbf{w}^{\top}\mathbf{x}_{i}+b\leq-1$
for all $(\mathbf{x}_{i},y_{i})\in D\,\wedge\, y_{i}=-1$ (or equivalently,
$y_{i}(\mathbf{w}^{\top}\mathbf{x}_{i}+b)\geq1$ for all $(\mathbf{x}_{i},y_{i})\in D$
if we combine the two inequations), as shown in Figure \ref{fig:svm}.
Simple calculation shows that the margin equals $\frac{2}{\left\Vert \mathbf{w}\right\Vert _{2}}$.
Therefore, to maximize the margin, we can instead minimize $\left\Vert \mathbf{w}\right\Vert _{2}$,
which leads to the objective below:

\begin{equation}
\begin{array}{c}
{\displaystyle \arg\min_{\mathbf{w},b,\mathbf{\xi}}\;\left\Vert \mathbf{w}\right\Vert _{2}^{2}}\\
\mbox{s.t. }y_{i}(\mathbf{w}^{\top}\mathbf{x}_{i}+b)\ge1\mbox{ for }i=1,2,\cdots,n.
\end{array}
\end{equation}
In practice, the training dataset $D$ usually contains instances
that are \emph{noises} or \emph{outliers} (i.e., instances with wrong
labels). To tolerate these instances, the SVM does not insist the
positive and negative instances to be placed exactly at the two sides
of the margins. It introduces a slack variable $\xi_{i}$, $\xi_{i}\geq0$,
for each instance $\mathbf{x}_{i}$ in $D$ and requires only $y_{i}(\mathbf{w}^{\top}\mathbf{x}_{i}+b)\ge1-\xi_{i}$.
So noises and outliers can be placed inside the margin or even at
opposite sides. This gives the objective of linear SVM:

\begin{equation}
\begin{array}{c}
{\displaystyle \arg\min_{\mathbf{w},b,\mathbf{\xi}}\;\left\Vert \mathbf{w}\right\Vert _{2}^{2}+C\sum_{i=1}^{n}\xi_{i}}\\
\mbox{s.t. }y_{i}(\mathbf{w}^{\top}\mathbf{x}_{i}+b)\ge1-\xi_{i}\mbox{ and }\\
\xi_{i}\ge0,\mbox{ for }i=1,2,\cdots,n.
\end{array}\label{eq:svm}
\end{equation}
Note that the slacks are minimized as well (in the term $\sum_{i=1}^{n}\xi_{i}$)
in order to penalize the noises and outliers and keep their numbers
small. The hyperparameter $C$ controls the trade-off between maximizing
the margin and minimizing the number of noises/outliers.

The linear SVM can be extended to nonlinear SVM by utilizing the \emph{kernel
trick}. Define a function $\phi:\mathbb{R}^{d}\rightarrow\mathbb{R}^{d'}$,
$d'>d$, that maps an instance $\mathbf{x}$ to a point in a higher
(possibly infinite) dimensional space, the nonlinear SVM finds a separating
hyperplane in that space. Since $d'>d$, the found hyperplane may
not be linear anymore in the original $d$-dimensional input space.
It can be shown that, if we choose $\phi$ carefully such that the
inner product $\phi(\mathbf{a})^{\top}\phi(\mathbf{b})$ can be represented
by a kernel function $K:\mathbb{R}^{d}\times\mathbb{R}^{d}\rightarrow\mathbb{R}$
(i.e., $\phi(\mathbf{a})^{\top}\phi(\mathbf{b})=K(\mathbf{a},\mathbf{b})$)
for any $\mathbf{a},\mathbf{b}\in\mathbb{R}^{d}$, then we can solve
the objective of nonlinear SVM in a manner whose complexity is \emph{independent
of the higher dimension} $d'$.  This is known as the kernel trick.
The nonlinear SVM usually makes better predictions than the linear
SVM does when the input dimension $d$ is not very high.

\textbf{Practical Considerations.} The objective of conventional SVM
(either linear or nonlinear) can be solved by standard quadratic programming
software. However, when applied to an SNS service like Facebook, the
solver needs to deal with an extremely large $D$ due to the huge
user base owned by the SNS service. To speed up the training process,
we adopt the \emph{Smooth SVM} (SSVM) \cite{lee2001ssvm} in this
paper. The SSVM, a variant of SVM, adds $\frac{b^{2}}{2}$ into the
objective of SVM and employs the squares of slacks $\xi_{i}^{2}$
to penalize the noises/outliers. The SSVM utilizes the KKT optimization
condition to convert the conventional SVM to an unconstrained minimization
problem that can be solved efficiently using the Newton's method
with an Armijo stepsize.

The kernel trick applies to the SSVM too. In our experiment, we pair
up the nonlinear SSVM with the \emph{RBF kernel}, which is defined
as $K(\mathbf{a},\mathbf{b})=e^{-\gamma\left\Vert \mathbf{a}-\mathbf{b}\right\Vert _{2}^{2}}.$

There are 2 hyperparameters we have to determine in the nonlinear
SSVM: the penalty coefficient $C$ and $\gamma$ in the RBF kernel
function. We use the uniform design model selection method \cite{UD}
with 9-13 stages to search for an appropriate combination of these
hyperparameters.

\subsection{Feature Selection}

\label{sec:detection-feature}The training of SSVM is preceded by
a \emph{feature selection process} \cite{wrapper}, where we select
only a subset of features in $D$ for the training. This process is
necessary because 1) given a tremendous amount of sessions (Figure
\ref{fig:detect-scheme}(1)) that will be monitored by the SNS servers,
it helps the SSVM \emph{scale up in making predictions} by considering
only a small set of features; 2) the selected features give us a hint
on what is useful to distinguish the stalkers from the account owners.
By ignoring those features that are not helpful, we can collect fewer
actions (Figure \ref{fig:detect-scheme}(2)) and save the cost of
feature extraction (Figure \ref{fig:detect-scheme}(3)) on each SNS
server; and 3) our results show that it improves the prediction accuracy
of the final SSVM we obtain.

\begin{figure}[t]
\includegraphics[width=\columnwidth]{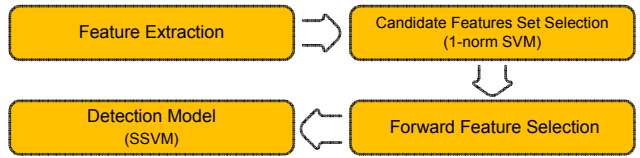}
\caption{\label{fig:model-dev} The steps in training a detection model.}
\end{figure}

The feature selection precess is divided into two stages, as shown
in Figure \ref{fig:model-dev}. In the first stage, we use the \emph{1-norm
SVM} \cite{onesvm} to determine a candidate set of features. In the
second stage, we use the \emph{forward feature selection} \cite{wrapper}
algorithm to determine the best final features from the candidate
set for training the detection model.

Unlike the SVM which minimizes $\left\Vert \mathbf{w}\right\Vert _{2}^{2}$
in its objective (Eq. (\ref{eq:svm})), the 1-norm SVM minimizes $\left\Vert \mathbf{w}\right\Vert _{1}^{2}$
(called the LASSO penalty \cite{lasso}) instead. We employ the 1-norm
SVM to determine the candidate set because it usually finds a sparse
$\mathbf{w}$ (i.e., $\mathbf{w}$ that tends to have zeros) thanks
to its ``compressed sensing'' interpretation \cite{CS}. We obtain
the candidate set by keeping only those features that correspond to
the non-zeros in $\mathbf{w}$, as the features corresponding to zeros
are usually redundant or noise features \cite{onesvm}.

And then, we use the forward feature selection algorithm to determine the final
features from the candidate set. The algorithm starts with an empty set for keeping the
final features. At each step, one feature from the candidate set that improves the
prediction accuracy\footnote{We use the 10-fold cross validation \cite{wrapper} to
measure the accuracy.} of SSVM most is added to this set. The algorithm then repeats
the above step until the candidate set becomes empty or there is no feature in the
candidate set that improves the accuracy.

\section{Performance Evaluation}

\label{sec:eval}

In this section, we evaluate the performance of the proposed detection model.

\subsection{Settings and Metrics}

After data cleaning (described in Section~\ref{sec-behavior-collect}), there are 278
instances (i.e., sessions) in $D$, among which 178 instances are positive (i.e.,
labeled by $+1$, which denotes ``acquaintance'' or ``stranger'') and 100 instances are
negative (i.e., labeled by $-1$, which denotes ``account owner''). Each instance is
presented by 139 feature values.

To the best of our knowledge, currently there is no other detection scheme for the
usage stealing problem in SNS services. Thus we compare the detection model to itself
by imposing different observation periods. Specifically, given an observation period
$L$, we extract the feature values of a session only from those actions that are
performed within $L$ minutes after the start of the session. We study the performance
of the detection model given $L=1,2,\cdots,25$ minutes. Although a subject was asked
browse an SNS account for 30 minutes during each round of the experiment described in
Section~\ref{sec-behavior-collect}, we set the maximal value of $L$ to 25 rather than
30 because some subjects appear to lose patience and become idle after 25 minutes.
Under the premise of data consistency, we consider $L\leq25$ here.

As described in Sections~\ref{sec:detection-feature} and
Section~\ref{sec:detection-model}, to obtain our detection model, we first employ the
1-norm SVM to get the candidate features, and then use the forward feature selection
and SSVM with 10-fold cross validation to find the best final features and best
combination of the hyperparameters $C$ and $\gamma$. We report the performance of our
model by using the leave-one-out
cross validation \cite{wrapper} on $D$%
\footnote{We do not partition $D$ into the \emph{training}, \emph{validation},
and \emph{testing sets} because $D$ is not large enough.%
}.

To measure the performance, we first obtain a confusion table shown in
Table~\ref{tab:confusion} by counting the numbers of \emph{True Positives} (TP),
\emph{False Positives} (FP), \emph{True Negatives} (TN), and \emph{False Negatives}
(FN) made by the model when predicting the labels of instances in the testing dataset.
\begin{table}
\begin{centering}
{\small{}}%
\begin{tabular}{cc|cc}
 &  & \multicolumn{2}{c}{{\small{Predicted}}}\tabularnewline
 &  & {\small{Pos.}} & {\small{Neg.}}\tabularnewline
\hline
\hline
\multirow{2}{*}{{\small{Real}}} & {\small{Pos.}} & {\small{\#True Positives (TP) }} & {\small{\#False Negatives (FN) }}\tabularnewline
 & {\small{Neg.}} & {\small{\#False Positives (FP) }} & {\small{\#True Negatives (TN) }}\tabularnewline
\end{tabular}
\par\end{centering}{\small \par}

\caption{\label{tab:confusion}The confusion matrix.}
\end{table}
And then we calculate the following metrics:

\begin{eqnarray*}
Accuracy\quad= & \frac{TP+TN}{TP+TN+FP+FN},\\
FPR\quad= & \frac{FP}{TN+FP},\\
FNR\quad= & \frac{FN}{TP+FN},\\
TPR\quad= & \frac{TP}{TP+FN},\\
Precision\quad= & \frac{TP}{TP+FP},\\
Recall\quad= & \frac{TP}{TP+FN},\\
F\mbox{-}score\quad= & 2*\frac{Precision*Recall}{Precision+Recall},
\end{eqnarray*}
with various observation periods.

\subsection{Detection Performance at 25 Minutes}

\begin{table}
\begin{centering}
{\small{}}%
\begin{tabular}{c|c|c}
 & \multicolumn{1}{c|}{{\small{With Oversampling}}} & \multicolumn{1}{c}{{\small{Without Oversampling}}}\tabularnewline
\hline
\hline
{\small{With }} & {\small{Acc: 93.53\%}} & {\small{Acc: 90.29\%}}\tabularnewline
{\small{Feature }} & {\small{FPR: 5.00\%}} & {\small{FPR: 18.00\%}}\tabularnewline
{\small{Selection}} & {\small{FNR: 7.30\%}} & {\small{FNR: 5.06\%}}\tabularnewline
\hline
{\small{$F$-score}} & {\small{0.9483}} & {\small{0.9260}}\tabularnewline
\hline
\hline
{\small{Without}} & {\small{Acc: 91.37\%}} & {\small{Acc: 87.77\%}}\tabularnewline
{\small{Feature }} & {\small{FPR: 6.00\%}} & {\small{FPR: 22.00\%}}\tabularnewline
{\small{Selection}} & {\small{FNR: 10.11\%}} & {\small{FNR: 6.74\%}}\tabularnewline
\hline
{\small{$F$-score}} & {\small{0.9302}} & {\small{0.9071}}\tabularnewline
\end{tabular}{\small{}}
\par\end{centering}{\small \par}

\caption{\label{tab:confusion-25}The results achieved under various conditions.}

\end{table}

\begin{figure}[t]
\includegraphics[scale=0.38]{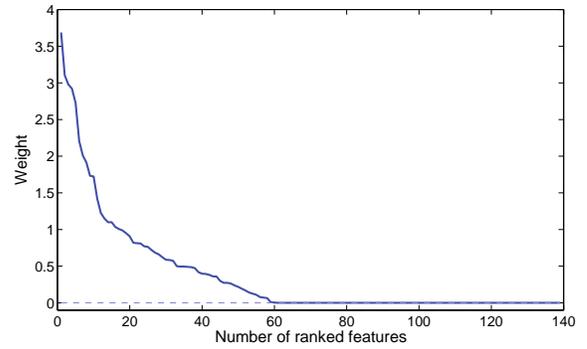} \caption{\label{fig:weight}The weights in $\mathbf{w}$ found by the 1-norm
SVM over the corresponding features. The features are ranked by their
weights. Only 60 features remain in the candidate feature set.}
\end{figure}

We first study the performance of our detection model when $L=25$ minutes.
Table~\ref{tab:confusion-25} shows the results achieved by the model \emph{with} and
\emph{without} feature selection. As we can see, feature selection does improve the
performance by giving higher accuracy/$F$-scores and lower FPR/FNR. This is because the
noisy features are successfully eliminated. Figure~\ref{fig:weight} shows that there
are only 60 features remain after the applying the 1-norm SVM for candidate set
selection.

\begin{figure}[t]
\includegraphics[scale=0.38]{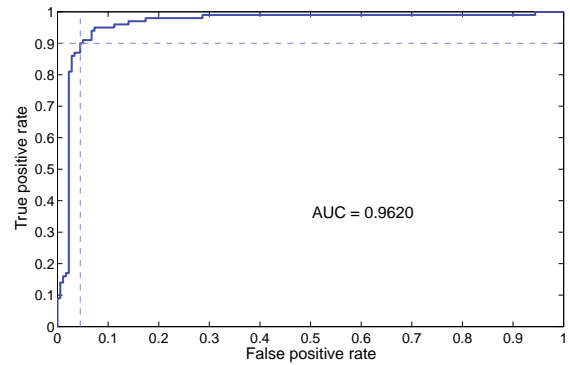} \caption{\label{fig:roc-25} The ROC curve and AUC given by the model at 25-minutes.}
\end{figure}

We notice that the dataset $D$ is imbalanced---the ratio of positive instances to
negative ones is $1.78:1$. Since there are more positive than negative instances, we
tend to obtain a higher FPR. To overcome this issue, we adopt an \emph{oversampling}
approach by randomly selecting and duplicating 78 negative instances to balance the
ratio between positive and negative instances. The effect of duplicating an instance is
to double the penalty if we misclassify the instance. So by duplicating negative
instances in $D$ we can avoid aliasing and reduce the FPR. Note that because there
exists randomness when applying the oversampling technique, we train 10 models and
average their results. Table~\ref{tab:confusion-25} shows the results achieved by our
model \emph{with} and \emph{without} oversampling. We can see that the oversampling
successfully controls the trade-off between FPR and FNR.

Figure \ref{fig:roc-25} shows the ROC curve and AUC of our model when both the feature
selection and oversampling are applied. We get a fairly high AUC (0.962). In
particular, the ROC curve shows that we can achieve 90\% TPR at 4.5\% FPR.

\subsection{Early Detection Performance}

\begin{figure}[t]
\includegraphics[scale=0.38]{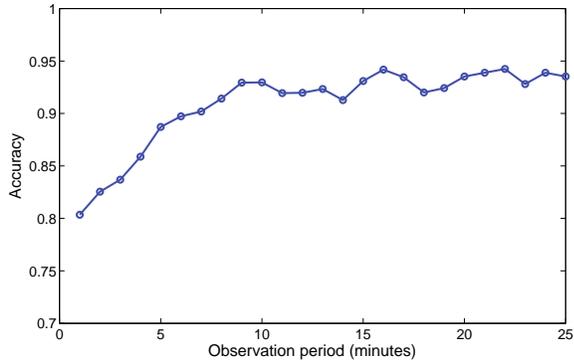} \caption{\label{fig:acc-time} Accuracy for every minute. It shows that our
detection model can achieve stable and reasonably good results after
7 minutes.}
\end{figure}

To prevent the leak of sensitive information, we should perform the usage stealing
detection as early as possible for each session. To see how our model performs with
time limits, we vary $L$ from 1 to 25 minutes and train respective models for each $L$
with feature selection and oversampling. Figure~\ref{fig:acc-time} shows the accuracy
achieved by these models. After 7 minutes, we can get stable and reasonably good
results, with the accuracy rate higher than 90\% with $L\geq 7$ minutes. Even at 2
minutes, we obtain an accuracy above 80\%, which is still satisfactory when the scheme
is used as a trigger for more sophisticated analysis.

\begin{figure}[t]
\includegraphics[scale=0.38]{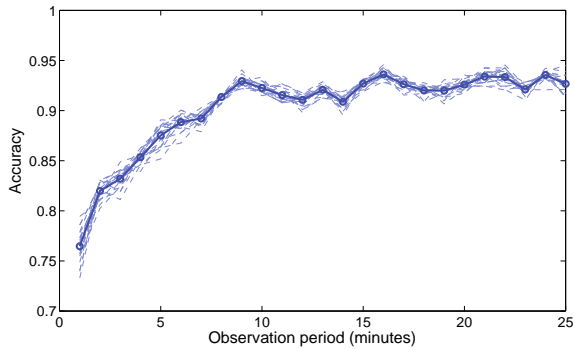}
\caption{\label{fig:acc-time-10f}The accuracy achieved by 20 models trained using the
10-fold cross validation on 20 randomly permuted datasets. The thick line represents
the average accuracy.}
\end{figure}

To test the robustness of our model, we randomly permute $D$ for 20 times and train one
model using the 10-fold cross validation~\cite{wrapper} for each of the 20
permutations. Figure~\ref{fig:acc-time-10f} and Table~\ref{tab:acc-time-std} show the
mean accuracy and standard deviation given by the 20 models. The results indicate that
the accuracy has a very low standard deviation regardless of $L$. In addition,
comparing Figure~\ref{fig:acc-time-10f} with Figure~\ref{fig:acc-time}, we can see that
our model performs consistently no matter it is trained (using the 10-fold cross
validation) or tested (using the leave-one-out cross validation), which means the
performance of our detection scheme is very robust.

\begin{table}
\begin{centering}
{\small{}}%
\begin{tabular}{c|cccccc}
{\small{Minute}} & {\small{2}} & {\small{3}} & {\small{4}} & {\small{5}} & {\small{6}} & {\small{7}}\tabularnewline
\hline
{\small{Mean}} & {\small{81.9\%}} & {\small{83.2\%}} & {\small{85.3\%}} & {\small{87.5\%}} & {\small{88.8\%}} & {\small{89.2\%}}\tabularnewline
{\small{Std.}} & {\small{0.7\%}} & {\small{0.8\%}} & {\small{0.5\%}} & {\small{1.0\%}} & {\small{0.8\%}} & {\small{0.5\%}}\tabularnewline
\hline
\hline
 & {\small{8}} & {\small{9}} & {\small{10}} & {\small{11}} & {\small{12}} & {\small{13}}\tabularnewline
\hline
{\small{Mean}} & {\small{91.3\%}} & {\small{92.9\%}} & {\small{92.2\%}} & {\small{91.5\%}} & {\small{91.0\%}} & {\small{92.0\%}}\tabularnewline
{\small{Std.}} & {\small{0.3\%}} & {\small{0.4\%}} & {\small{0.4\%}} & {\small{0.6\%}} & {\small{0.5\%}} & {\small{0.4\%}}\tabularnewline
\hline
\hline
 & {\small{14}} & {\small{15}} & {\small{16}} & {\small{17}} & {\small{18}} & {\small{19}}\tabularnewline
\hline
{\small{Mean}} & {\small{90.8\%}} & {\small{92.7\%}} & {\small{93.6\%}} & {\small{92.6\%}} & {\small{92.0\%}} & {\small{92.0\%}}\tabularnewline
{\small{Std.}} & {\small{0.5\%}} & {\small{0.3\%}} & {\small{0.4\%}} & {\small{0.5\%}} & {\small{0.4\%}} & {\small{0.6\%}}\tabularnewline
\hline
\hline
 & {\small{20}} & {\small{21}} & {\small{22}} & {\small{23}} & {\small{24}} & {\small{25}}\tabularnewline
\hline
{\small{Mean}} & {\small{92.5\%}} & {\small{93.3\%}} & {\small{93.3\%}} & {\small{92.1\%}} & {\small{93.5\%}} & {\small{92.6\%}}\tabularnewline
{\small{Std.}} & {\small{0.4\%}} & {\small{0.5\%}} & {\small{0.7\%}} & {\small{0.4\%}} & {\small{0.4\%}} & {\small{0.8\%}}\tabularnewline
\end{tabular}
\par\end{centering}{\small \par}

\caption{\label{tab:acc-time-std}Mean and standard deviation of the accuracy given by
the models trained using the 10-fold cross validation on 20 randomly permuted
datasets.}
\end{table}

\section{Security Analysis}

\label{sec:security}

As shown in Figure~\ref{fig:detect-scheme}, all the data collection, processing,
decision, and follow-up actions (such as challenges and punishment) in our scheme can
all be performed on the server side. So there is no way for attackers to compromise the
scheme from the clients.

Since our detection methodology is running at the server side (i.e., at operators), the
attackers cannot evade our detection scheme---once logged-in, each user (including the
attacker) must be monitored by an SNS server running our scheme. The only way for an
attacker to continuously use the victim's account is to evade the detection model.

The detection model does not rely on any cryptography technology and is completely
user-behavior-based. So, in order not to be detected by the model, attackers have to 1)
mimic the owners\textquoteright{} actions; or 2) do as few actions as possible and run
away. The attackers of the first kind are less likely because the
owner\textquoteright{}s action model is not well known~\cite{joinson:looking:08}. Even
if some attackers read this paper and successfully mimic the owners, they are forced to
spend time on something they are not really interested and skip some information they
desire more. This makes the attacks less harmful. For the attackers of the second kind,
our scheme imposes a high time pressure because the detection model can achieve close
to 80\% accuracy even if an attackers browse the victims' newsfeeds for only 1 minute.
Again, the time pressure makes the attacks less harmful because the attackers may not
be able to find the information they want within such a limited time.

Note that our detection scheme is not tied to any specific detection model. For
example, a personalized detection model can be particularly helpful to identify the
attackers of the first kind because it is even harder to imitate each individual's
behavior. Also, a detection model that takes the timestamp of each action into account
may be helpful to identify the attackers of the second kind, because users (either the
account owners or stalkers) often take actions in some order they are used to. In fact,
while this work firstly points out a new direction for future research against the
usage stealing, it is certainly possible to develop more sophisticated detection models
to fight against the ever-smarter attackers.

\section{Conclusion}

\label{sec:concl}

In this paper, we have proposed a novel continuous authentication approach for SNS that
analyzes users' browsing behavior to detect usage stealing incidents. We use Facebook
as a case study and show that 1) the \emph{role-driven behavioral diversity} does
exist; 2) based on the so-called role-driven behavioral diversity property, we can
design a low-cost detection scheme applicable to \emph{all} users; and 3) the scheme is
hard to evade and it renders reasonable detection performance after an observation
period of 2 minutes.

\begin{figure}[t]
\includegraphics[width=\columnwidth]{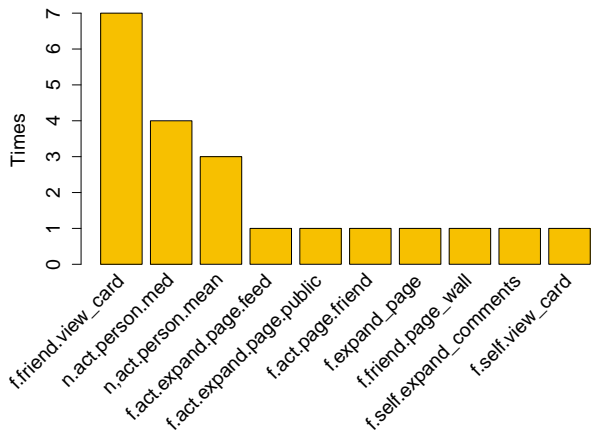}
\caption{\label{fig:top-feature-stalker}Top-3 features used to identify the
positive instances (stalkers) in the first 7 minutes.}
\end{figure}

\begin{figure}[t]
\includegraphics[width=\columnwidth]{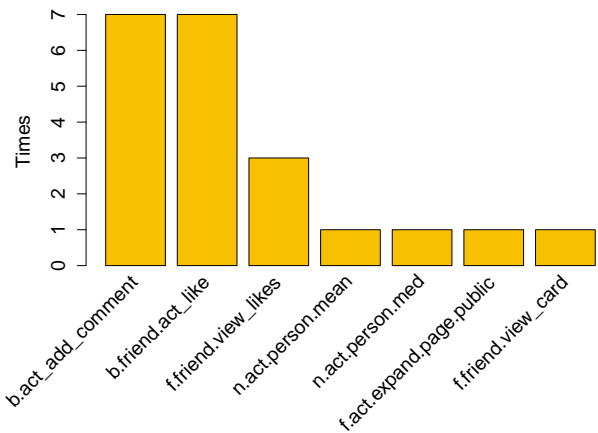} \caption{\label{fig:top-feature-owner}Top-3 features used to ientify the negative
instances (owners) in the first 7 minutes.}
\end{figure}

As future work, we plan to study the browsing behavior of individuals and develop
personalized detection models. These models can be triggered only when needed and serve
as the detailed analyzers for suspicious sessions. We also plan to improve our low-cost
detection model to give higher detection accuracy within the first 7 minutes. Such an
improvement is possible because we see different user behavior in short- and long-term.
To share our observation, we count features corresponding to the 3 most positive and 3
most negative weights in $\mathbf{w}$ identified by SSVM when the observation period
$L$ varies from 1 to 7. Figures \ref{fig:top-feature-stalker} and
\ref{fig:top-feature-owner} show the histograms of counts of the 3 most positively- and
negatively-weighted features respectively. Some features are rather surprising as they
are not prominent in the full 30-minute traces discussed in
Section~\ref{sec:behavior-diversity}. We hope this study can motivate in-depth studies
on developing more sophisticated models against usage stealing issues.

\bibliographystyle{plain}
\bibliography{ref}

\end{document}